\documentstyle[emulateapj,apjfonts,psfig]{article}

\begin{document}

\title{\large \bf The Mean Number of Extra Micro-Image Pairs for Macro-Lensed Quasars}

\author{Jonathan Granot\altaffilmark{1},
Paul L. Schechter\altaffilmark{2,1} and 
Joachim Wambsganss\altaffilmark{3}}

\altaffiltext{1}{Institute for Advanced Study, Olden Lane, Princeton, NJ 08540}
\altaffiltext{2}{Department of Physics, Massachusetts Institute of Technology,
77 Massachusetts Avenue, Cambridge, MA 02139}
\altaffiltext{3}{Universit\"at Potsdam, Institut f\"ur Physik, 
Am Neuen Palais 10, 14467 Potsdam, Germany}

\begin{abstract}

When a gravitationally lensed source crosses a caustic, a pair of images is 
created or destroyed. We calculate the mean number of such pairs of micro-images 
$\langle n\rangle$ for a given macro-image of a gravitationally lensed point 
source, due to microlensing by the stars of the lensing galaxy. This quantity was 
calculated by Wambsganss, Witt \& Schneider (1992) for the case of zero external 
shear, $\gamma=0$, at the location of the macro-image. Since in realistic lens 
models a non-zero shear is expected to be induced by the lensing galaxy, 
we extend this calculation to a general value of $\gamma$. We find a complex
behavior of $\langle n\rangle$ as a function of $\gamma$ and the normalized 
surface mass density in stars $\kappa_*$. Specifically, we find that at high 
magnifications, where the average total magnification of the macro-image is 
$\langle\mu\rangle=|(1-\kappa_*)^2-\gamma^2|^{-1}\gg 1$, $\langle n\rangle$ 
becomes correspondingly large, and is proportional to $\langle\mu\rangle$. 
The ratio $\langle n\rangle/\langle\mu\rangle$ is largest near the line 
$\gamma=1-\kappa_*$ where the magnification $\langle\mu\rangle$ becomes infinite, 
and its maximal value is 0.306. We compare our semi-analytic results for 
$\langle n\rangle$ to the results of numerical simulations and find good agreement.
We find that the probability distribution for the number of extra micro-image pairs is 
reasonably described by a Poisson distribution with a mean value of $\langle n\rangle$,
and that the width of the macro-image magnification distribution tends to be largest 
for $\langle n\rangle\sim 1$.
\end{abstract}

\keywords{cosmology: gravitational lensing, dark matter---quasars: general}

\section{Introduction}
\label{sec:intro}

Gravitational microlensing by the stars of a lensing galaxy can have
a large effect on the magnification of lensed sources (Chang \& Refsdal 
1979; Young 1981; Paczy\'nski 1986). As the macro-images of multiply imaged 
sources are typically located in relatively dense star fields of the lensing 
galaxy, microlensing is quite common in such systems. The typical angular 
separation between the micro-images of a cosmological source due to a
stellar mass micro-lens is of order $1$ micro-arc-second, which is far too small to be 
resolved. For the time being, the only observable manifestation of microlensing is 
to change the magnification of the macro-image relative to the average magnification 
that is predicted for the smoothed out surface mass density profile of the galaxy. 

The first observational evidence for quasar microlensing 
was found by Irwin et al. (1989) in the quadruple system
Q2237+0305, which subsequently  has been monitored by many groups
(Corrigan et al. 1991;  Burud et al. 1998; Lewis et al. 1998,
Wo{\'z}niak et al. 2000a,b). In particular the latest results show  
that all four quasar images vary dramatically, going up and down  
by more than one magnitude on timescales of less than a year.
The fact that individual (caustic-crossing) events can be clearly
distinguished allows to put upper limits on the source  size
(Wambsganss, Paczy\'nski \& Schneider 1990; Yonehara 1999, 2001;
Wyithe et al. 2000).

In the double quasar Q0957+561,  originally there was an almost  linear 
change detected in the (time-shifted) brightness ratio between the two 
images ($ \Delta m_{\rm AB} \approx 0.25\,{\rm mag}$ over 5 years), which was
interpreted as microlensing by solar type stars. But since about 1991, 
this ratio stayed more or less ``constant" within about 0.05 mag, so 
not much microlensing was going on in this system recently
(Schild 1996; Pelt et al. 1998). However, even the ``lack of microlensing''
in this system can be used to put limits on compact dark matter in the
halo of the lensing galaxy (Wambsganss et al. 2000).  

A number of other multiple 
quasar systems are being monitored more or less
regularly, with some showing indications of microlensing, e.g.
H1413+117 ({\O}stensen et al. 1997),
B0218+357 (Jackson et al. 2000), or
HE1104-1805 (Gil-Merino et al. 2002, Schechter et al. 2002).
For a recent review  on quasar microlensing see Wambsganss (2001).

Microlensing has recently been suggested as the source of short time scale 
low-level variability in the time delayed light curves of multiple images of 
quasars (Wyithe \& Loeb 2002), and might help study the properties of broad-line 
clouds in quasars. A better understanding of the microlensing by stars in galaxies 
would provide a better handle on the probability distribution for microlensing 
of cosmological sources (Wyithe \& Turner 2002) that is relevant for Gamma-Ray 
Bursts (GRBs), especially in light of a possible microlensing event that was 
observed in the optical and near IR light curve of the afterglow of GRB 000301C 
(Garnavich, Loeb \& Stanek 2000; Gaudi, Granot \& Loeb 2001; 
Koopmans \& Wambsganss 2001)

Another example where microlensing may play an important role is in explaining 
the flux ratio anomalies observed in close pairs of images of quadruply lensed 
quasars (Schechter \& Wambsganss 2002). Such systems are usually modeled using 
a simple smooth surface mass 
density profile for the galaxy, possibly with the addition of an external shear. 
While these models successfully reproduce the observed locations of the 
macro-images, the flux ratios they predict are quite often in poor agreement 
with observations. Specifically, the theoretical flux ratio for a close pair of 
highly magnified macro-images is 1:1, while observations show a difference of up 
to one magnitude. For example, MG0414+0534 has an observed flux ratio of 2:1 in the 
optical (Hewitt et al. 1992; Schechter \& Moore 1993), while in the radio the flux
ratio is 1:1 (Trotter et al. 2000). There are also alternative explanations for 
these flux ratio anomalies, such as intervening dust (Lawrence et al. 1995) or 
milli-lensing by galactic sub-structure (Mao \& Schneider 1998; Metcalf \& Madau 
2001; Dalal \& Kochanek 2002; Chiba 2002). The study of microlensing can help distinguish 
between these alternative explanations, and may be useful in constraining the 
sub-structure of galaxies.

In many cases the study of microlensing cannot be done analytically, and much of 
the work is done using numerical simulations. The magnification distributions of the 
macro-image (MDMs) are particularly important for understanding the observed 
properties of lensed systems. The latter have been calculated for a wide range of 
parameters using numerical simulations (Wambsganss 1992; Rauch et al. 1992; Lewis 
\& Irwin 1995, 1996; Schechter \& Wambsganss 2002). While simulations are 
applicable to a wide range of problems, and are in many cases the only available 
technique, they are usually time consuming and do not always provide a qualitative
understanding of the results. Specifically, they do not seem to provide an 
explanation for the detailed structure that is present in the MDMs 
that they produce. 

In a recent paper, Schechter and Wambsganss (2002) explained the flux ratio 
anomalies as resulting from a different qualitative behavior of the MDMs for 
macro-minima and macro-saddlepoints in the arrival time surface,
that results in a larger probability for de-magnification (relative to the 
average magnification) of saddlepoints, compared to minima. This is in agreement 
with the observations of MG0414+0534 and four other recently discovered quadruple 
systems (Reimers et al. 2002; Inada et al. 2002; Burles et al, 
in preparation; Schechter et al., in preparation) in all of which the fainter 
image in a pair corresponds to a saddlepoint, while the brighter image
corresponds to a minimum. 

The structure of the MDMs seems to be tightly related to the probability 
distribution for the number of extra micro-image pairs (EIPs) corresponding to a given 
macro-image (Rauch et al. 1992). This may be expected since the magnification of the 
macro-image is simply the sum of the magnifications of all the micro-images that it is 
composed of. An analytic expression for the mean number of 
EIPs was derived by Wambsganss, Witt \& Schneider (1992, hereafter WWS) 
for the case of zero external shear. In \S \ref{AR} we generalize this result for
arbitrary values of the external shear, and obtain a semi-analytic expression.
More detailed expressions are provided in Appendix A, along with an analytic result
for the case of equal shear and convergence in stars. In \S \ref{num} we compare our 
analytic results to numerical simulations, and find good agreement. 
We also show that the probability distribution for the number of extra micro-image pairs is 
reasonably described by a Poisson distribution. 
The possible implications of our results are discussed in \S \ref{discussion}.

\section{Analytic Results}
\label{AR}

In this section we calculate the mean number of positive parity micro-images 
$\langle N\rangle$, and the mean number of extra micro-image pairs (EIPs) 
$\langle n\rangle$, that are induced by the random star field of the lensing 
galaxy near the location of a macro-image. The macro images are located at 
stationary points (i.e. minima, saddlepoints or maxima) of the time delay 
(Fermat) surface of the smoothed out surface mass density distribution of the 
lensing galaxy. The source size is assumed to be small compared to the Einstein 
radius of the stars in the galaxy and compared to the typical distance between 
caustics, so that we may assume a point source. In this limit
the probability distribution of the random shear caused by the star field and 
the resulting distributions of the number of micro-images and total magnification 
of the macro-image are independent of the mass spectrum of the stars (Schneider 
\& Weiss 1988). The remaining parameters on which the microlensing characteristics
in this case might still depend are the normalized surface mass density in stars 
$\kappa_*$ and in a smooth component (dark matter) $\kappa_c$, as well as the 
large scale shear of the galaxy $\gamma$. However, Paczy\'nski (1986) has shown 
that such a model is always equivalent to a model with effective convergence 
and shear given by
\begin{equation}
\kappa_{*}^{eff}={\kappa_{*}\over(1-\kappa_c)}\quad,\quad
\gamma^{eff}={\gamma\over(1-\kappa_c)}\ ,
\end{equation}
and with no smooth component ($\kappa_c^{eff}\equiv 0$). The effective magnification in 
this model is related to the true magnification by $\mu^{eff}=\mu(1-\kappa_c)^2$. 
Therefore, without loss of generality, we restrict ourselves to $\kappa_c=0$,
while letting $\kappa_*$ and $\gamma$ vary.

Each pair of micro-images (EIP) consists of a saddlepoint and a minimum, and thereby 
includes one positive parity image. For a macro-minimum there is also an 
additional positive parity micro-image corresponding to the global minimum. 
Therefore, $n=N-1$ for a macro-minimum and $n=N$ otherwise, where $n$ is 
the number of EIPs, and $N$ is the number of positive parity 
micro-images. This also carries through to the average values of these quantities:
\begin{equation}\label{n}
\langle n\rangle=\left\{\matrix{\langle N\rangle-1 & \ \ \kappa_*+\gamma<1 
\cr & \cr \langle N\rangle & \ \ \kappa_*+\gamma>1}\right.\ .
\end{equation}
It is therefore sufficient to calculate one of these quantities in order to 
determine the other. The values of $\langle N\rangle$ and $\langle n\rangle$ 
for the case of zero external shear ($\gamma=0$) were calculated by WWS. In this 
work we follow their analysis and generalize their result to the case of a 
non-zero shear $\gamma$. 
The average total magnification of the macro-image is given by
\begin{equation}\label{mu_av}
\langle\mu\rangle={1\over|(1-\kappa_*)^2-\gamma^2|}\ .
\end{equation}
As the magnification is solely due to area distortion, flux conservation implies 
that a sufficiently large area ${\mathcal A}_d$ in the deflector plane will be 
(backward) mapped onto an area ${\mathcal A}_s={\mathcal A}_d/\langle\mu\rangle$ in the 
source plane (when projected onto the deflector plane, so that it extends the same 
solid angle from the observer). 
The mean number of positive parity micro-images $\langle N\rangle$ is equal to 
the multiplicity $q$ by which regions of positive parity within ${\mathcal A}_d$, 
when mapped onto ${\mathcal A}_s$, cover ${\mathcal A}_s$.

\begin{figure*}
\centerline{\hbox{\psfig{figure=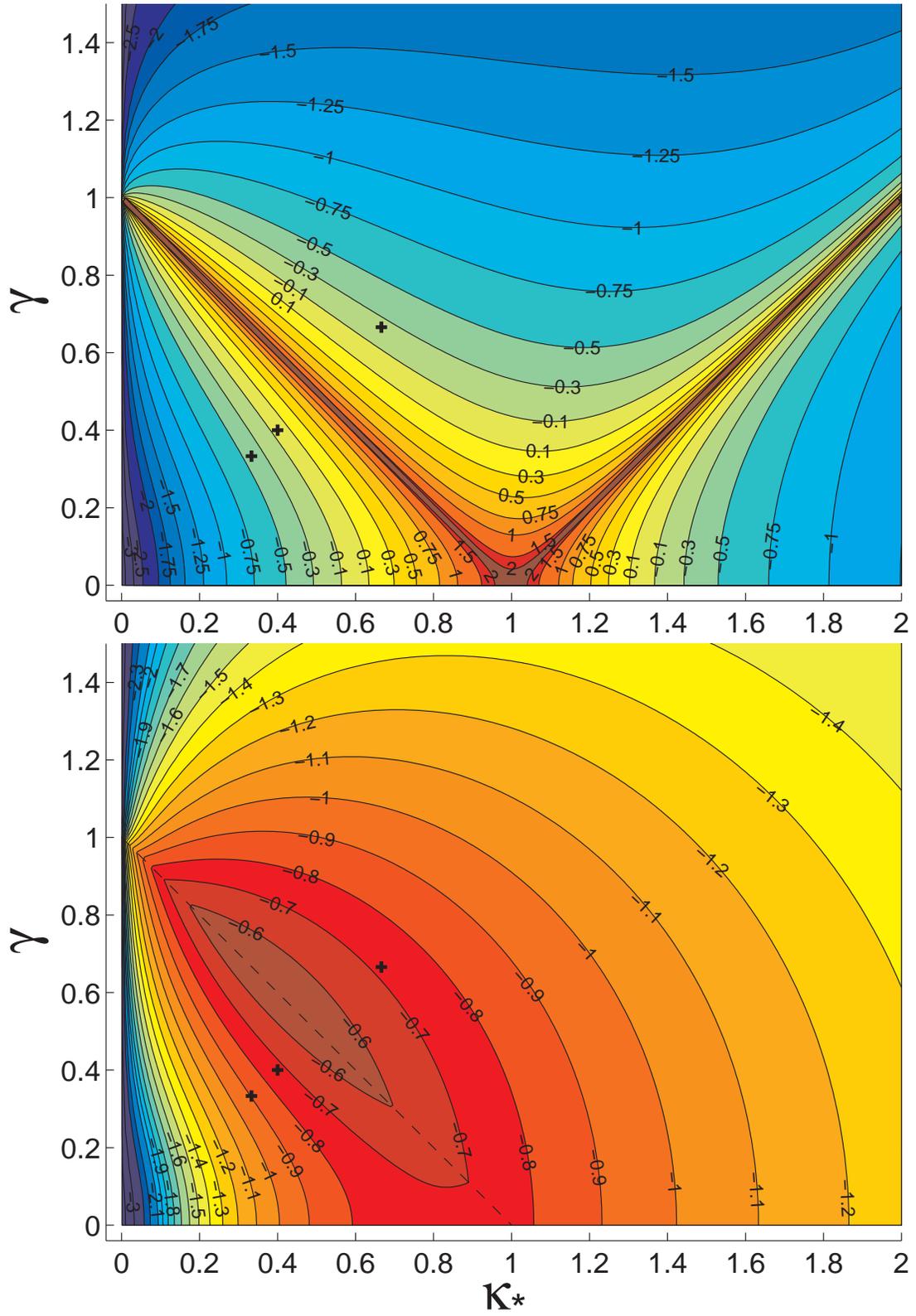,width=15cm}}}
\caption[]{\label{contour_plots} 
Contour plots of $\log_{10}(\langle n\rangle)$ (upper panel)
and $\log_{10}(\langle n\rangle/\langle\mu\rangle)$ (lower panel)
in the $\kappa_{*}-\;\gamma$ plane. The dashed line in the lower 
panel shows the line of infinite magnification $\gamma=1-\kappa_{*}$
along which there is a ``ridge'' in $\langle n\rangle/\langle\mu\rangle$, 
while the three plus symbols in both panels mark the values used in our 
three simulations: $\kappa_*=\gamma=0.333,\,0.400,\,0.666$.}
\end{figure*}

The probability distribution of the random shear produced by the stars is given by
\begin{equation}\label{P_s}
p(\kappa_*,S_1,S_2)={\kappa_*\over 2\pi(\kappa_*^2+S_1^2+S_2^2)^{3/2}}\ .
\end{equation}
(Nityananda \& Ostriker 1984; Schneider, Ehlers \& Falco 1992) where $S_1$ and 
$S_2$ are the two components of the {\em internal} shear. As we assume that $\kappa_c=0$ 
and that the stars are point masses, the convergence vanishes everywhere (except at the 
locations of the stars, where it is infinite, but these form a set of measure 
zero for any finite area on the deflector plane) and the local magnification at 
a given location in the star field is just due to the shear at that point
\begin{equation}\label{mu}
\mu(\gamma,S_1,S_2)={1\over 1-(\gamma+S_1)^2-S_2^2}\ ,
\end{equation}
where for convenience we have chosen $S_1$ to lie in the direction of the 
external shear $\gamma$. The area $da_d={\mathcal A}_d
p(\kappa_*,S_1,S_2)dS_1dS_2$ within ${\mathcal A}_d$ where the shear lies between
$(\gamma+S_1,S_2)$ and $(\gamma+S_1+dS_1,S_2+dS_2)$, is mapped onto the area 
$da_s=da_d/\mu(\gamma,S_1,S_2)$. Therefore
\begin{equation}\label{N_av}
\langle N\rangle=q={1\over{\mathcal A}_s}\int_{\mu>0}{da_d\over\mu}=
\langle\mu\rangle
\int\int_{\mu>0}{p(\kappa_*,S_1,S_2)dS_1dS_2\over\mu(\gamma,S_1,S_2)}=
\end{equation}
$$
{\kappa_*\over 2\pi|(1-\kappa_*)^2-\gamma^2|}\int_{-1-\gamma}^{1-\gamma}dS_1
\int_{-\sqrt{1-(\gamma+S_1)^2}}^{\sqrt{1-(\gamma+S_1)^2}}dS_2\,
{1-(\gamma+S_1)^2-S_2^2\over(\kappa_*^2+S_1^2+S_2^2)^{3/2}}\ .
$$
This integral may be evaluated analytically. However, the resulting
expression is long and cumbersome so that we prefer not to write it down 
explicitly. Instead, we show contour plots of $\langle n\rangle$ and 
$\langle n\rangle/\langle\mu\rangle$ in Figure \ref{contour_plots}, and provide the 
values of $\langle N\rangle/\langle\mu\rangle$ at representative points of 
$(\kappa_*,\gamma)$ in Table \ref{table1}. In Appendix A we reduce the expression 
in equation (\ref{N_av}) to a one dimensional integral in two different ways (so 
that $\langle N\rangle$ may be easily evaluated numerically) and provide an 
analytic expression for the case where $\gamma=\kappa_*$. For $\gamma=0$, equation 
(\ref{N_av}) reduces to a simple analytic expression (WWS):
\begin{equation}\label{N_av(gamma=0)}
{\langle N\rangle\over\langle\mu\rangle}=
\langle N\rangle(1-\kappa_*)^2=
1+2\kappa_*^2-2\kappa_*\sqrt{1+\kappa_*^2}\ .
\end{equation}
For $\kappa_*=0$, there are no extra images due to stars, and we have either
one positive parity image for a macro-minimum ($N=\langle N\rangle=1$ for 
$\gamma<1$) or none for a macro-saddlepoint ($N=\langle N\rangle=0$ for 
$\gamma>1$). Hence, for $\kappa_*=0$ we have:
\begin{equation}\label{N_av(kappa_*=0)}
{\langle N\rangle\over\langle\mu\rangle}=
\langle N\rangle(1-\gamma^2)=\left\{\matrix{1-\gamma^2 & \ \ \gamma<1 
\cr & \cr 0 & \ \ \gamma\geq 1}\right.\ .
\end{equation}

As can be seen from equation (\ref{N_av}), the ratio 
$\langle N\rangle/\langle\mu\rangle$ remains finite and varies smoothly
with $\gamma$ and $\kappa_*$ near the lines of infinite average total magnification 
$\langle\mu\rangle$ in the $\gamma-\kappa_*$ plane (i.e. $\gamma=|1-\kappa_*|$). 
Together with equation (\ref{n}) this implies the same for the ratio $\langle n
\rangle/\langle\mu\rangle$ along the line $\gamma=\kappa_*-1$, while along the line
$\gamma=1-\kappa_*$ it is continuous but its derivative is discontinuous in any 
direction that is not along this line (as can be seen in the lower panel of figure 
\ref{contour_plots}). 

Furthermore, $\langle n\rangle/\langle\mu\rangle$
decreases on either side of the line $\gamma=1-\kappa_*$, and therefore attains its
maximal value along this line, $(\langle n\rangle/\langle\mu\rangle)_{\rm max}=
0.306136$ at $(\kappa_*,\gamma)=(0.37895,0.62105)$. For $\langle\mu\rangle\gg 1$
($\gamma\approx|1-\kappa_*|$), small changes in $\gamma$ or $\kappa_*$ can cause 
large changes in $\langle\mu\rangle$ and $\langle n\rangle$, while the ratios 
$\langle n\rangle/\langle\mu\rangle$ or $\langle N\rangle/\langle\mu\rangle$ 
remain approximately constant. In this region, when crossing the line of infinite 
magnification $\gamma=1-\kappa_*$, from a macro-minimum ($\gamma<1-\kappa_*$) to a 
macro- saddlepoint ($\gamma>1-\kappa_*$), $\langle N\rangle$ decreases by $1$ 
(corresponding to the macro-minimum that disappears), but since $\langle N\rangle$ 
is infinite at this line, this amounts to a zero fractional change in 
$\langle N\rangle$.

\section{Comparison to Numerical Simulations}
\label{num}

\begin{figure*}
\centerline{\hbox{\psfig{figure=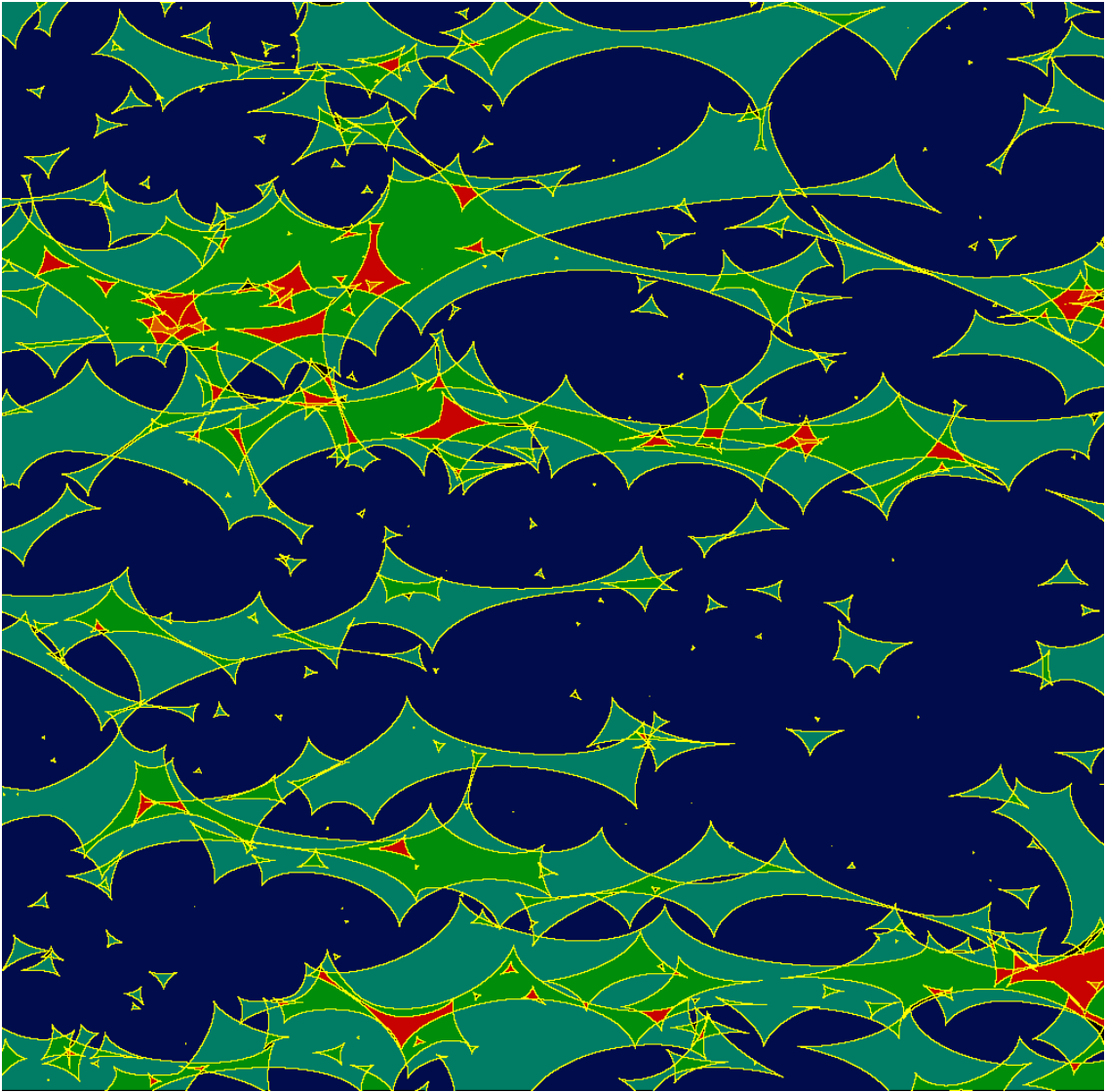,width=15cm}}}
\caption[]{\label{n_fig} 
A map of the number of extra micro-image pairs $n$ in the source plane,
for the simulation with $\gamma=\kappa_*=0.666$. The regions in dark blue,
turquoise, green, red and orange, correspond to $n=0,\,1,\,2,\,3$ and $4$,
respectively. The yellow lines represent the caustics, while regions 
for which $n$ could not be determined are in black.}
\end{figure*}

In this section we compare the analytic results of \S \ref{AR} to the results 
of numerical simulations. Combining a ray shooting code (Wambsganss 1990, 1999) 
with a program that detects the location of the caustics (Witt 1993) as in WWS,
we extract additional information on image multiplicity and magnification. 
Whenever the source crosses a caustic, 
a pair of images consisting of a micro-saddlepoint and a micro-minimum are created or 
destroyed. Each simulation is based on a particular realization of the random star 
field, that determines a caustic network in the source plane, and in principle 
determines the number of such extra image pairs (EIPs) at any point in the source 
plane. In order to calculate the average number of EIPs, $\langle n\rangle$, from 
the results of a simulation, we identify the regions with different number of EIPs, 
$n$, on the source plane, and calculate the fraction $f_n$ of the source plane that 
they cover, which is equal to the probability $p_n$ of having $n$ EIPs at a random 
location on the source plane. This is illustrated in Figure \ref{n_fig}.
Pixels in the source plane that are crossed by caustics (which are colored in yellow 
in Figure \ref{n_fig}) are attributed to the corresponding higher image number. 
There are also parts of the source plane for which it is very difficult to
uniquely identify $n$ (corresponding to the black regions in Figure \ref{n_fig}) 
due to the occasionally complex caustic structure, combined with the finite pixel size
of the simulation. These regions are left out when 
calculating $f_n$. The   ``unidentified'' 
regions correspond to $\sim 1\%$ of the source plane, and a large fraction of 
these regions probably corresponds to relatively high values of $n$. However, as
we do not know exactly what distribution of multiplicity $n$ we should assign to 
these regions, and lacking a better option, we simply assign to them the same 
distribution of the identified regions:
\begin{equation}\label{p_n}
p_n={f_n\over\sum_{n'}f_{n'}}\ .
\end{equation}
This is equivalent to leaving out the unidentified regions entirely, and normalizing 
the probability distribution of $p_n$.
The average number of pairs for the simulation is then given by
\begin{equation}\label{N_av_sim}
\langle n\rangle_{\rm sim}=\sum_{n}n\, p_n\ .
\end{equation}

Since we have simple analytic expressions for $\langle n\rangle$ for $\gamma=0$
(WWS, see equation \ref{N_av(gamma=0)}) and for $\kappa_*=0$ (equation 
\ref{N_av(kappa_*=0)}) we chose values of $(\kappa_*,\gamma)$ along the line 
$\kappa_*=\gamma$ for the simulations, so that they would serve as a good check for 
our analytic results. An additional advantage of this choice is that it corresponds to
the interesting case of a singular isothermal sphere (for $\kappa_c=0$).
We performed three simulations, two for macro-minima ($\kappa_*=\gamma=0.333,\,0.400$ 
with $\langle\mu\rangle\cong 3,\,5$) and one for a macro-saddlepoint 
($\kappa_*=\gamma=0.666$ with $\langle\mu\rangle\cong 3$). 
The results of the simulations are shown in Tables \ref{table2} and \ref{table3}.
For $\kappa_*=\gamma=0.333,\,0.400$ and $0.666$, $\langle n\rangle_{\rm sim}$ is 
$3.7\%$, $3.3\%$ and $7.7\%$, respectively, lower than the analytic result 
$\langle n\rangle$ from \S \ref{AR} (denoted by $\langle n\rangle_{\rm th}$ in
Table \ref{table2}). The values of $\langle n\rangle_{\rm sim}$ in Table \ref{table2} 
were calculated according to equation (\ref{N_av_sim}) which assigns the $p_n$ 
distribution of the regions with identified image multiplicity $n$ in the source 
plane to the unidentified regions. However, as mentioned above, the unidentified regions 
are typically related to complicated caustic structures, and are hence more likely to
contribute to larger values of $n$ compared to the identified regions. If the average 
$n$ of the unidentified regions is say, 5, this would make $\langle n\rangle_{\rm sim}$
$1.1\%$ higher, $2.6\%$ higher and $1.1\%$ lower than $\langle n\rangle$,
for $\kappa_*=\gamma=0.333,\,0.400$ and $0.666$, respectively.
The latter should assume values of 3.96, 3.19 and 5.70, for $\kappa_*=\gamma=0.333,
\,0.400$ and $0.666$, respectively, in order for $\langle n\rangle_{\rm sim}$ to be 
exactly equal to $\langle n\rangle$. One should also keep in mind
the ``cosmic'' variance, i.e. the fluctuations between the results of different 
simulations for the same $\kappa_*$ and $\gamma$, due to different statistical 
realizations of the star field over a finite region in the deflector plane. 
We therefore conclude that there is good agreement between the 
results of the numerical simulations and our analytic results for the average number
of EIPs $\langle n\rangle$.

The average total magnifications from the simulations, $\langle\mu\rangle_{\rm sim}$,
are $3.1\%$, $0.16\%$ and $3.5\%$ lower than their theoretical values for 
$\kappa_*=\gamma=0.333,\,0.400$ and $0.666$, respectively. The scatter in 
$\langle\mu\rangle_{\rm sim}$ can be attributed to the ``cosmic'' variance.
The fact that $\langle\mu\rangle_{\rm sim}$ is on average slightly lower than its
theoretical value arises since we consider finite regions in the deflector plane and in
the source plane. Rays that fall very close to a star in our deflector field suffer 
very large deflection angles which may take them outside of our source field; these 
are not compensated for by rays with large deflection angles from stars outside of 
our deflector field, that should have been deflected into our source field
(see Katz, Balbus, Paczy\'nski 1986; Schneider \& Weiss 1987).
We conclude that the numerical simulations are in good agreement with the theory
on the value of $\langle\mu\rangle$, as well.

The magnification distribution of the macro-image (MDM), $p(\mu)$, where $p(\mu)d\mu$ is 
the probability that the total magnification of the macro-image is between $\mu$ and $\mu+d\mu$,
can be expressed as a sum over the contributions from regions with different numbers of EIPs $n$,
\begin{equation}\label{p(mu)}
p(\mu)=\sum_{n=0}^\infty p_n(\mu)\ ,
\end{equation}
where $p_n(\mu)d\mu=(dp_n/d\mu)d\mu$ is the probability of having $n$ EIPs and a total
magnification between $\mu$ and $\mu+d\mu$, with the normalization
\begin{equation}\label{norm_p_n}
\int_0^\infty p_n(\mu)d\mu=p_n\quad,\quad \int_0^\infty p(\mu)d\mu=1\ .
\end{equation}

\begin{figure*}
\centerline{\hbox{\psfig{figure=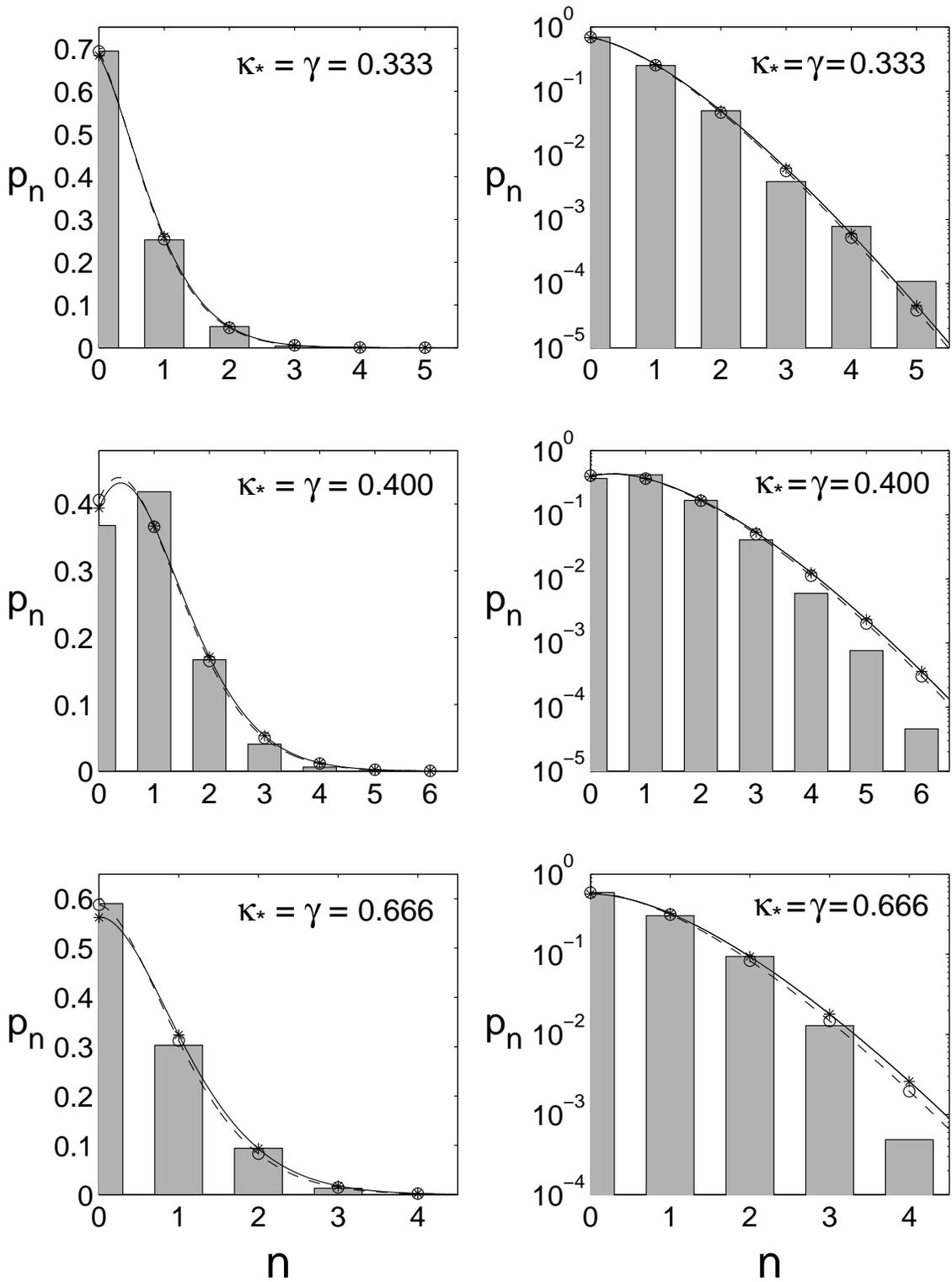,width=15cm}}}
\caption[]{\label{p_n_fig} 
The bars show the probability $p_n$ of having $n$ extra image pairs, calculated 
from our simulations ($\kappa_*=\gamma=0.333,\,0.400,\,0.666$), using equation 
(\ref{p_n}). The lines correspond to a Poisson distribution [equation (\ref{Poisson})] 
where the solid line is for the analytical (theoretical) mean value 
$\langle n\rangle$, and the dashed line is for the mean value from the
simulation $\langle n\rangle_{\rm sim}$, calculated according to equation 
(\ref{N_av_sim}). The theoretical values of $p_n$ for integer values of $n$ are shown
by the asterisk and circle symbols. The panels on the left (right) column are 
linear (logarithmic) representations of $p_n$.}
\end{figure*}

In Figure \ref{p_n_fig} we show the results for $p_n$ from our simulations
(that are given in Table \ref{table2}), along with a Poisson 
distribution:
\begin{equation}\label{Poisson}
p_n={e^{-a}a^n\over n!}\ ,
\end{equation}
where the solid line is for $a=\langle n\rangle$ and the dashed line is for
$a=\langle n\rangle_{\rm sim}$, calculated according to equation (\ref{N_av_sim}).
A Poisson distribution provides a reasonable fit to the results of all our 
simulations. For large values of $n$ there is a relatively larger deviation
from a Poisson distribution. This results, in part, from the difficulty in identifying 
regions with large $n$ in the source plane. One should also keep in mind that 
the ``cosmic'' variance in $p_n$ becomes larger with increasing $n$. There still seem 
to be some systematic deviations from a Poisson distribution, however they appear to be 
small for relatively low values of $n$, which cover most of the source plane. We therefore
consider a Poisson distribution for $p_n$ to be a reasonable approximation.

\begin{figure*}
\centerline{\hbox{\psfig{figure=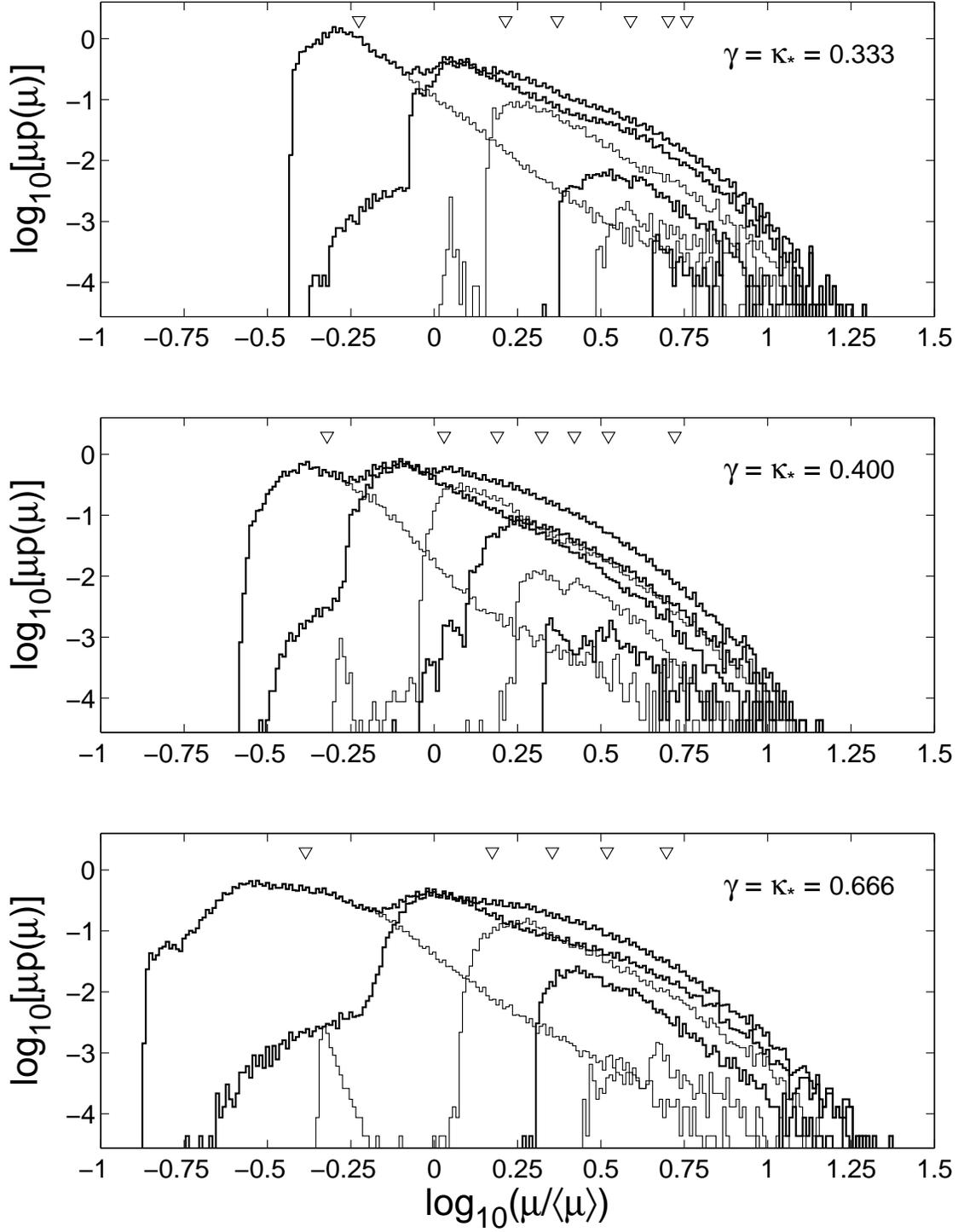,width=15cm}}}
\caption[]{\label{hist} 
A log-log plot of the magnification distribution for the macro-image $\mu p(\mu)$ 
(uppermost line), for our simulations, and its decomposition, $\mu p_n(\mu)$, to the 
contributions from regions with different numbers of extra image pairs, $n$. 
The alternating thin and thick solid lines, from left to right, are for $n=0$ through $6$. 
The inverted triangles represent the average magnification from regions with $n$ extra 
image pairs, $\langle\mu\rangle_n$.
}
\end{figure*}

Figure \ref{hist} shows plots of $\mu p(\mu)$ and $\mu p_n(\mu)$ from our 
simulations. The shape of $p_n(\mu)$ for different $n$ seems quite similar,
while the average magnification,
\begin{equation}\label{mu_n}
\langle\mu\rangle_n={1\over p_n}\int_0^\infty p_n(\mu)\mu d\mu\ ,
\end{equation}
and the normalization, $p_n$, are different.

\section{Discussion}
\label{discussion}

We have calculated the mean number of extra micro-image pairs $\langle n\rangle$
of a point source, as a function of $\kappa_*$ and $\gamma$. The results are shown 
in Figure \ref{contour_plots} and Table \ref{table1}. One dimensional integrals for 
general values of $(\kappa_*,\gamma)$ and an analytic result for the case 
$\gamma=\kappa_*$ are presented in Appendix A. Near the lines of infinite magnification 
in the $\gamma-\kappa_*$ plane ($\gamma=|1-\kappa_*|$), $\langle n\rangle$ diverges, 
and is proportional to the mean macro-magnification $\langle\mu\rangle$. The ratio 
$\langle n\rangle/\langle\mu\rangle$ is continuous along these lines and varies smoothly
along the line $\gamma=\kappa_*-1$, while its derivative is discontinuous along the 
line $\gamma=1-\kappa_*$ in directions that  are not along this line. This creates a 
``ridge'' in $\langle n\rangle/\langle\mu\rangle$ along the line $\gamma=1-\kappa_*$,
where it also peaks at $(\kappa_*,\gamma)=(0.379,0.621)$ with $(\langle n\rangle
/\langle\mu\rangle)_{\rm max}=0.306$. The analytic results for $\langle n\rangle$ 
are in good agreement with the results of numerical simulations we performed for 
$\kappa_*=\gamma=0.333,\,0.400$ and $0.666$ (as can be seen in Table \ref{table2}). 

We find that the probability distribution $p_n$ for the number of extra image pairs 
(EIPs) $n$, that is calculated from numerical simulations, may be reasonably described by a 
Poisson distribution. This result holds both for the numerical simulations performed 
in this paper (e.g. Table \ref{table3} and Figures \ref{n_fig} and \ref{p_n_fig}),
and for numerical simulations from previous works (Rauch et al. 1992). Furthermore, 
the shape of the magnification distribution $p_n(\mu)$ of regions with a given $n$ 
appears to be similar for different values of $n$, where only the overall normalization 
$p_n$ and mean magnification $\langle\mu\rangle_n$ depend on $n$ (see Figure \ref{hist}).

The mean number of EIPs $\langle n\rangle$ can serve as a rough 
measure for the width of $p(\mu)$, the magnification distributions of the 
macro-image (MDM). For $\langle n\rangle\ll 1$ there is little contribution to the 
MDM from regions in the source plane with $n>0$, since these regions cover only a 
small fraction of the source plane. For $\langle n\rangle\gg 1$ the Poisson 
distribution $p_n$, for $n$, approaches a Gaussian distribution with a mean value 
of $\langle n\rangle$ and a standard deviation of $\sigma=\langle n\rangle^{1/2}$, 
so that only $\sim\langle n\rangle^{1/2}$ different values of $n$, around 
$n\approx\langle n\rangle$ will have a noticeable contribution to the MDM. We also 
note that the average magnification from regions with $n$ EIPs, $\langle\mu\rangle_n$, 
is approximately linear in $n$ (see Table \ref{table3}), so that we expect 
$\Delta\mu/\langle\mu\rangle\sim\langle n\rangle^{-1/2} \ll 1$.
Therefore, the width of the MDM is expected to be largest for 
$\langle n\rangle\sim 1$. This seems to be in rough agreement with the results of 
numerical simulations (Wambsganss 1992; Irwin \& Lewis 1995; 
Schechter \& Wambsganss 2002).

\acknowledgements

This research was supported by grants NSF PHY 00-70928 (JG) and NSF AST 02-06010 (PLS).
PLS is grateful to the John Simon Guggenheim Foundation for the award of a fellowship.
We thank Hans Witt for providing his code for the determination of the location of 
the caustics, and Scott Gaudi for his careful reading of the manuscript and useful comments.

\section{Appendix A}

The inner integral in equation (\ref{N_av}) may be solved analytically, 
reducing it to a single integral
\begin{equation}\label{N_av1}
{\langle N\rangle\over\langle\mu\rangle}={\kappa_*\over 2\pi}
\int_{-1-\gamma}^{1-\gamma}dS_1\left[\ln\left({A-B\over A+B}\right)
+{2AB\over (\kappa_*^2+S_1^2)}\right]\ ,
\end{equation}
where $\langle\mu\rangle=|(1-\kappa_*)^2-\gamma^2|^{-1}$ [see equation 
(\ref{mu_av})] and
\begin{equation}\label{N_av1a}
A\equiv\sqrt{1+\kappa_*^2-\gamma(2S_1+\gamma)}\quad,\quad 
B\equiv\sqrt{1-(S_1+\gamma)^2}\ .
\end{equation}
This integral may be easily evaluated numerically. Alternatively, changing 
variables in equation (\ref{N_av}) to $r$ and $\phi$ where $x=\gamma+S_1$, 
$y=S_2$, $r^2=x^2+y^2$ and $\tan(\phi)=y/x$, gives
\begin{equation}\label{N_av2}
{\langle N\rangle\over\langle\mu\rangle}={\kappa_*\over 2\pi}
\int_0^1dr\int_0^{2\pi}d\phi\,{r(1-r^2)\over(\kappa_*^2+\gamma^2+r^2
-2\gamma r\cos\phi)^{3/2}}
\end{equation}
$$
={2\kappa_*\over\pi}\int_0^1dr\,{r(1-r^2)\,{\rm EllipticE}
\left({\pi\over 2},{-4\gamma r\over \kappa_*^2+(r-\gamma)^2}\right)
\over\left[\kappa_*^2+(r+\gamma)^2\right]\sqrt{\kappa_*^2+(r-\gamma)^2}}\ ,
$$
where
\begin{eqnarray}\label{EllipticE}
{\rm EllipticE}(\phi,x)&\equiv&\int_0^{\phi}d\theta\,(1-x\sin^2\theta)^{1/2}\ .
\end{eqnarray}

For the case of $\gamma=\kappa_*$, corresponding to a singular isothermal sphere,
we obtain an analytic result:
\begin{eqnarray}\label{N_av_SIS}
{\langle N\rangle\over\langle\mu\rangle}={1\over\sqrt{\gamma}\,\pi}
\left\{ {\over}\hspace{-0.14cm}  
-6\gamma^2{\rm EllipticE}\left[{\rm arcsin}(C),C^{-2}\right]
\right.\quad\quad\quad\quad\quad\quad\ \nonumber\\
-\,3(1-\gamma)\gamma^2{\rm EllipticF}\left[{\rm arcsin}(C),C^{-2}\right]
\quad\quad\quad\quad\nonumber\\
+\,C\left[\gamma(2\gamma^2-1){\rm EllipticK}(C^2)\quad\quad\quad\quad\right.
\quad\quad\quad\ \ \nonumber\\
\quad\quad\quad\left.+\,D\,\,{\rm EllipticPi}(E,C^2)
+D^*{\rm EllipticPi}(E^*,C^2)\right]
\left.{\over}\hspace{-0.2cm}\right\}\ ,
\end{eqnarray}
where
\begin{eqnarray}\label{CDE}
C\equiv\sqrt{4\gamma\over 1+2\gamma+2\gamma^2}\quad,\quad
D\equiv{i+(1-i)\gamma+(2-i)\gamma^2-(1+3i)\gamma^3\over 4}\nonumber\ ,\\
E\equiv{2\over 1+(1+i)\gamma}\ ,\quad\quad\quad\quad\quad\quad\quad\quad
\end{eqnarray}
$D^*$ ($E^*$) is the complex conjugate of $D$ ($E$), and
\begin{eqnarray}\label{Elliptic_FKPI}
{\rm EllipticF}(\phi,x)\equiv\int_0^{\phi}d\theta\,(1-x\sin^2\theta)^{-1/2}\ ,
\quad\quad\quad\quad\quad\quad\ \, 
\nonumber\\
{\rm EllipticK}(x)\equiv{\rm EllipticF}(\pi/2,x)\ ,\quad\quad\quad\quad
\quad\quad\quad\quad\quad\ \, \ \nonumber\\
{\rm EllipticPi}(x,y)\equiv\int_0^{\pi/2}d\theta\,
(1-x\sin^2\theta)^{-1}(1-y\sin^2\theta)^{-1/2}\ .\,
\end{eqnarray}


\newcommand{\rb}[1]{\raisebox{1.5ex}[0pt]{#1}}
\begin{deluxetable}{ccllllllllllll}
\tablecaption{The ratio $\langle N\rangle/\langle\mu\rangle$.\label{table1}}
  \tablehead{ \colhead{$\gamma$} & 
  \colhead{$\kappa_*$=\,0}\hspace{-0.2cm} & \colhead{0.1} & \colhead{0.2} & \colhead{0.3} & 
  \colhead{0.4} & \colhead{0.5} & \colhead{0.6} & \colhead{0.7} & 
  \colhead{0.8} & \colhead{0.9} & \colhead{1} & \colhead{1.2} & \colhead{1.5}}
  \startdata 0 & 1 & 0.8190 & 0.6721 & 0.5536 & 0.4584 & 
  0.3820 & 0.3206 & 0.2711 & 0.2310 & 0.1983 & 0.1716 & 0.1311 & 0.0917 \\ \hline
  0.1 & 0.99 & 0.8105 & 0.6650 & 0.5478 & 0.4537 & 0.3782 & 
  0.3176 & 0.2687 & 0.2292 & 0.1969 & 0.1704 & 0.1303 & 0.0913 \\ \hline
  0.2 & 0.96 & 0.7850 & 0.6438 & 0.5305 & 0.4398 & 0.3672 &   
  0.3089 & 0.2618 & 0.2237 & 0.1926 & 0.1670 & 0.1281 & 0.0901 \\ \hline
  0.3 & 0.91 & 0.7428 & 0.6089 & 0.5022 & 0.4171 & 0.3492 &
  0.2947 & 0.2507 & 0.2149 & 0.1856 & 0.1615 & 0.1246 & 0.0882 \\ \hline
  0.4 & 0.84 & 0.6841 & 0.5608 & 0.4635 & 0.3864 & 0.3250 &
  0.2756 & 0.2357 & 0.2031 & 0.1763 & 0.1540 & 0.1198 & 0.0856 \\ \hline
  0.5 & 0.75 & 0.6095 & 0.5005 & 0.4156 & 0.3487 & 0.2955 &
  0.2526 & 0.2176 & 0.1889 & 0.1650 & 0.1450 & 0.1140 & 0.0824 \\ \hline
  0.6 & 0.64 & 0.5200 & 0.4296 & 0.3602 & 0.3058 & 0.2621 &
  0.2266 & 0.1973 & 0.1728 & 0.1522 & 0.1348 & 0.1073 & 0.0787 \\ \hline
  0.7 & 0.51 & 0.4173 & 0.3505 & 0.2999 & 0.2596 & 0.2266 &
  0.1989 & 0.1756 & 0.1557 & 0.1386 & 0.1239 & 0.1001 & 0.0746 \\ \hline
  0.8 & 0.36 & 0.3048 & 0.2678 & 0.2383 & 0.2131 & 0.1908 &
  0.1711 & 0.1536 & 0.1382 & 0.1245 & 0.1125 & 0.0925 & 0.0702 \\ \hline
  0.9 & 0.19 & 0.1907 & 0.1886 & 0.1805 & 0.1692 & 0.1568 &
  0.1444 & 0.1323 & 0.1211 & 0.1107 & 0.1012 & 0.0848 & 0.0657 \\ \hline
  1 & 0 & 0.0974 & 0.1236 & 0.1314 & 0.1310 & 0.1265 & 
  0.1200 & 0.1126 & 0.1049 & 0.0974 & 0.0902 & 0.0771 & 0.0611 \\ \hline
  1.2 & 0 & 0.0303 & 0.0534 & 0.0682 & 0.0764 & 0.0801 &
  0.0808 & 0.0795 & 0.0770 & 0.0739 & 0.0703 & 0.0628 & 0.0521 \\ \hline
  1.5 & 0 & 0.0111 & 0.0212 & 0.0298 & 0.0366 & 0.0415 & 
  0.0448 & 0.0468 & 0.0477 & 0.0478 & 0.0472 & 0.0449 & 0.0400 \\ \hline
 \enddata
  \tablecomments{The ratio of the mean number of positive parity micro-images 
$\langle N\rangle$ and the mean total magnification $\langle\mu\rangle$ for 
different values of the normalized surface mass density in stars, $\kappa_*$, 
and the external shear, $\gamma$. $\gamma$ and $\kappa_*$, as determined with equations 
(\ref{N_av}), (\ref{N_av1}), (\ref{N_av2}), (\ref{N_av_SIS}).}
\end{deluxetable}  
 
\begin{deluxetable}{ccccc}
\tablewidth{220pt}  
\tablecaption{Numerical Simulations Versus Theory\label{table2}}
\tablehead{\colhead{$\kappa_*=\gamma$} & 
\colhead{$\langle n\rangle_{\rm sim}$} & \colhead{$\langle n\rangle_{\rm th}$} &
\colhead{$\langle \mu\rangle_{\rm sim}$} & \colhead{$\langle \mu\rangle_{\rm th}$}}
  \startdata
0.333 & 0.367 & 0.3806 & 2.902 & 2.994 \\
0.400 & 0.901 & 0.9320 & 4.992 & 5.000 \\
0.666 & 0.531 & 0.5759 & 2.907 & 3.012 \\ \hline
\enddata
\tablecomments{The first column labels the simulation, through the values of
the normalized surface mass density in stars, $\kappa_*$, and the external shear,
$\gamma$. The next two columns provide the average number of extra image 
pairs calculated from the simulation (according to equation \ref{N_av_sim}), 
$\langle n\rangle_{\rm sim}$, and the theoretical prediction for this quantity,
$\langle n\rangle_{\rm th}$, calculated using equations (\ref{n}) and (\ref{N_av}). 
The last two columns are the average magnification
of the simulation $\langle \mu\rangle_{\rm sim}$, and the theoretical average 
magnification $\langle \mu\rangle_{\rm th}$ (from equation \ref{mu_av}).}
\end{deluxetable}  
 
\begin{deluxetable}{clllll}
\tablewidth{260pt}  
\tablecaption{Results of Numerical Simulations\label{table3}}
\tablehead{\colhead{$\kappa_*=\gamma$} & \colhead{$n$} & 
\colhead{$f_n$} & \colhead{$\sum_{i=0}^{n}f_i$} & \colhead{$p_n$} &
\colhead{$\langle \mu\rangle_n$}}
  \startdata
           & 0 & 0.690800 & 0.690800 & 0.693511 & 1.780 \\ 
           & 1 & 0.251230 & 0.942030 & 0.252216 & 4.903 \\
           & 2 & 0.049298 & 0.991328 & 0.049492 & 6.999 \\
\rb{0.333} & 3 & 0.003878 & 0.995206 & 0.003893 & 11.62 \\
           & 4 & 0.000777 & 0.995983 & 0.000780 & 15.07 \\
           & 5 & 0.000108 & 0.996091 & 0.000108 & 17.14 \\ \hline        
\vspace{-0.3cm} & & & & &  \\ 
       & 0 & 0.362692 & 0.362692 & 0.367695 & 2.387 \\ 
       & 1 & 0.412717 & 0.775409 & 0.418409 & 5.352 \\
       & 2 & 0.164497 & 0.939906 & 0.166766 & 7.726 \\
 0.400 & 3 & 0.039863 & 0.979769 & 0.040413 & 10.51 \\
       & 4 & 0.005832 & 0.985601 & 0.005912 & 13.16 \\
       & 5 & 0.000749 & 0.986350 & 0.000759 & 16.65 \\
       & 6 & 0.000045 & 0.986395 & 0.000046 & 26.33 \\ \hline        
\vspace{-0.3cm} & & & & & \\ 
       & 0 & 0.584765 & 0.584765 & 0.589852 & 1.239 \\ 
       & 1 & 0.300253 & 0.885018 & 0.302865 & 4.497 \\ 
 0.666 & 2 & 0.093123 & 0.978141 & 0.093933 & 6.807 \\
       & 3 & 0.012753 & 0.990894 & 0.012864 & 9.935 \\ 
       & 4 & 0.000482 & 0.991376 & 0.000486 & 14.90 \\ \hline
 \enddata
  \tablecomments{The first column labels the simulation, through the values of
$\kappa_*=\gamma$. The remaining five columns are the number of extra image pairs $n$,
the corresponding fraction of the source plane $f_n$, its cumulative value up to $n$,
its normalized value $p_n$ calculated according to equation (\ref{p_n}), and the mean 
magnification of regions with $n$ extra micro-image pairs, $\langle\mu\rangle_n$
(that are marked by inverted triangles in Figure \ref{hist}).}
\end{deluxetable}

\begin{thebibliography}{}

\bibitem[Burud et al. 1998]{B98}
Burud, I., Stabell, R., Magain, P., et al. 1998, A\&A, 339, 701

\bibitem[Chang \& Refsdal 1979]{CR79}
Chang, K., \& Refsdal, S. 1979, Nature, 282, 561

\bibitem[Chiba 2002]{C02}
Chiba, M. 2002, ApJ, 565, 17

\bibitem[Corrigan et al. 1991]{C91}
Corrigan, R.T., et al. 1991, AJ, 102, 34

\bibitem[Dalal \& Kochanek 2002]{DK02}
Dalal, N., \& Kochanek, C.S. 2002, ApJ, 572, 25

\bibitem[Gaudi, Granot \& Loeb 2001]{pGGL01} 
Gaudi, B.S., Granot, J., \& Loeb, A. 2001, 561, 178

\bibitem[Garnavich, Loeb \& Stanek 2000]{GLS00}
Garnavich, P.M., Loeb, A., \& Stanek, K.Z. 2000, ApJ, 544, L11

\bibitem[Merino, Wisotzki \& Wambsganss 2002]{MWW02}
Gil-Merino, R., Wisotzki, L., \& Wambsganss, J. 2002, A\&A, 381, 428

\bibitem[Hewitt et al. 1992]{Hewitt92}
Hewitt, J.N., Turner, E.L., Lawrence, C.R., 
Schneider, D.P, Brody, J.P. 1992, AJ, 104, 968

\bibitem[Inada et al. 2002]{Inada02}
Inada, N., et al. 2002, submitted to ApJ

\bibitem[Irwin et al. 1989]{I98}
Irwin, M.J., Webster, R.L., Hewett, P.C., Corrigan, R.T., Jedrzejewski, R.I. 1989,
AJ, 98, 1989

\bibitem[Jackson, Xanthopoulos \& Browne 2000]{JXB00}
Jackson, N.J., Xanthopoulos, E., \& Browne, I.W.A. 2000, MNRAS, 311, 389

\bibitem[Katz et al. 1986]{Katz86}
Katz, N., Balbus, S., \& Paczy\'nski, B. 1986, ApJ, 306, 2 

\bibitem[Koopmans \& Wambsganss 2001]{KW01}
Koopmans, L.V.E., \& Wambsganss, J. 2001, MNRAS, 325, 1317

\bibitem[Lawrence et al. 1995]{Lawrence95}
Lawrence, C.R., Elston, R., Januzi, B.T., 
\& Turner, E.L. 1995, AJ, 110, 2570

\bibitem[Lewis \& Irwin 1995]{LI95}
Lewis, G.F., \& Irwin M.J. 1995, MNRAS, 276, 103

\bibitem[Lewis \& Irwin 1996]{LI99}
Lewis, G.F., \& Irwin M.J. 1996, MNRAS, 283, 225

\bibitem[Lewis et al. 1998]{Lewis98}
Lewis, G.F., Irwin, M.J., Hewett, P.C., \& Foltz, C.B. 1998, MNRAS, 295, 573

\bibitem[Mao \& Scneider 1998]{MS98}
Mao, S., \& Scneider, P. 1998, MNRAS, 295, 587

\bibitem[Metcalf \& Madau 2001]{MM01}
Metcalf, R.B, \& Madau, P. 2001, ApJ, 563, 9

\bibitem[Nityananda \& Ostriker 1984]{NO84}
Nityananda, R., \& Ostriker, J.P. 1984, J. Astrophys. Astr., 5, 235

\bibitem[{\O}stensen 1997]{O97}
{\O}stensen, R., Remy, M., Lindblad, P.O., Refsdal, S., 
	Stabell, R. et al. 1997, A\&A (Suppl.), 126, 393

\bibitem[Paczy\'nski 1986]{P86}
Paczy\'nski, B. 1986, ApJ, 301, 503

\bibitem[Pelt et al. 1998]{Pelt98}
Pelt, J., Schild, R., Refsdal, S., \& Stabell, R. 1998, A\&A ,336, 829

\bibitem[Rauch et al. 1992]{Rauch92}
Rauch, K.P., Mao, S., Wambsganss, J., \& Paczy\'nski, B. 1992, ApJ, 386, 30

\bibitem[Reimers et al. 2002]{Reimers02}
Reimers, D., Hagen, H.J., Baade, R., Lopez, S., \&
Tytler, T. 2002, A\&A, 382, L26

\bibitem[Schechter \& Moore 1993]{SM93}
Schechter, P.L., \& Moore, C.B. 1993, AJ, 105, 1

\bibitem[Schechter \& Wambsganss (2002)]{SW02}
Schechter, P.L., \& Wambsganss, J. 2002, ApJ in press (astro-ph/0204425)

\bibitem[Schechter et al. 2002]{S02}
Schechter, P.L., Udalski, A., Szymanski, M., Kubiak, M., 
	Pietrzynski, G.  Soszynski, I.  Wo\'zniak, P.  Zebrun, K.  
	Szewczyk, O., \& Wyrzykowski, L. 2002, preprint (astro-ph/0206263)

\bibitem[Schild 1996]{Schild96}
Schild, R. 1996, ApJ, 464, 125


\bibitem[Schneider, Ehlers \& Falco 1992]{SEF92}
Schneider, P., Ehlers, J., \& Falco, E.E. 1992, Gravitational Lenses 
(Berlin: Springer), 329

\bibitem[Schneider \& Weiss 1987]{SW87}
Schneider, P., \& Weiss, A. 1987, A\&A, 171, 49

\bibitem[Schneider \& Weiss 1988]{SW88}
Schneider, P., \& Weiss, A. 1988, ApJ, 327, 526

\bibitem[Trotter et al. 2000]{Trotter}
Trotter, C.S., Winn, J.N. \& Hewitt, J.N. 2000, ApJ, 535, 671

\bibitem[Wambsganss 1990]{W90}
Wambsganss, J. 1990, PhD thesis, available as report MPA 550

\bibitem[Wambsganss 1992]{W92}
Wambsganss, J. 1992, ApJ, 386, 19

\bibitem[Wambsganss 1999]{W99}
Wambsganss, J. 1999, Journ. Comp. and Appl. Math., 109, 353

\bibitem[Wambsganss 2001]{W01}
Wambsganss, J. 2001, Publ. Astron. Soc. Austr., 18, 207

\bibitem[Wambsganss, Paczy{\'n}ski \& Schneider 1992]{WPS92}
Wambsganss, J., Paczy{\'n}ski, B., \& Schneider, P. 1990, ApJ, 358, L33

\bibitem[Wambsganss, Witt \& Schneider 1992]{WWS92}
Wambsganss, J., Witt, H.J., \& Schneider, P. 1992, A\&A, 258, 591 (WWS)

\bibitem[Wambsganss et al. 2000]{W00}
Wambsganss, J., Schmidt, R., Colley, W., Kundi\'c, T., \& Turner, E.L. 2000,  
	A\&A, 362, L37

\bibitem[Witt 1993]{W93}
Witt, H.J. 1993, ApJ, 403, 530


\bibitem[Wo\'zniak et al. 2000a]{Wozniak00a}
Wo{\'z}niak, P.R., Alard, C., Udalski, A., Szyma{\'n}ski, M., 
Kubiak, M., Pietrzy{\'n}ski, G., \& Zebru{\'n}, K. 2000a, ApJ, 529, 88

\bibitem[Wo\'zniak et al. 2000b]{Wozniak00ab}
Wo{\'z}niak, P.R., Udalski, A., Szyma{\'n}ski, M., Kubiak, M., 
Pietrzy{\'n}ski, G., Soszy{\'n}ski, I., \& Zebru{\'n}, K. 2000b, ApJ, 540, L65

\bibitem[Wyithe et al. 2000]{Wyithe00}
Wyithe, J.S.B,  Webster, R.L., Turner, E.L., \& Mortlock, D.J. 2000, 
	MNRAS, 315, 62 

\bibitem[Wyithe \& Loeb 2002]{WL02}
Wyithe, J.S.B., \& Loeb, A. 2002, Submitted to ApJ (astro-ph/0204529)

\bibitem[Wyithe \& Turner 2002]{WT02}
Wyithe, J.S.B., \& Turner, E.L. 2002, ApJ, 567, 18

\bibitem[Yonehara 1999]{Yonehara99}
Yonehara, A. 1999, ApJ, 519, L31

\bibitem[Yonehara 2001]{Yonehara01}
Yonehara, A. 2001, ApJ, 548, L127 

\bibitem[Young 1981]{Young81}
Young, P. 1981, ApJ, 244, 756

\end{thebibliography}
\end{document}